\documentclass[prd,preprint,eqsecnum,nofootinbib,amsmath,amssymb,preprintnumbers,tightenlines,longbibliography,floatfix,superscriptaddress]{revtex4-1}

\usepackage[mathscr]{eucal}
\usepackage{mathrsfs}
\usepackage{hyperref}
\usepackage{graphicx}
\usepackage{bm}

\usepackage{graphicx}
\usepackage{bm}

\def\x{{\mathbf x}}
\def\y{{\mathbf y}}

\def\z{{\mathbf z}}

\def\v{{\bm v}}

\def\p{{\bm{p}}}
\def\k{\bm k}

\def\v{{\bm v}}

\def\Im{{\rm Im}}

\def\y{{\bm y}}
\def\z{{\bm z}}

\def\x{{\bm x}}

\def\st{\begin{equation}}
\def\stp{\end{equation}}
\def\bg{\begin{eqnarray}}
\def\nd{\end{eqnarray}}
\def\Eq#1{Eq.~(\ref{#1})}

\def\Fig#1{Fig.~\ref{#1}}

\def\Eq#1{Eq.~(\ref{#1})}

\def\Fig#1{Fig.~\ref{#1}}

\def\lsim{\mbox{~{\protect\raisebox{0.4ex}{$<$}}\hspace{-1.1em}
	{\protect\raisebox{-0.6ex}{$\sim$}}~}}

\def\nott#1{\setbox0=\hbox{$#1$}                
   \dimen0=\wd0                                
   \setbox1=\hbox{/} \dimen1=\wd1               
   \ifdim\dimen0>\dimen1                        
      \rlap{\hbox to \dimen0{\hfil/\hfil}}      
      #1                                        
   \else                                        
      \rlap{\hbox to \dimen1{\hfil$#1$\hfil}}   
      /                                         
   \fi}                                         %

\advance\parskip 1.9pt
\advance\voffset -0.2in

\begin{document}
\title{Anomalous Viscosity of the Quark-Gluon Plasma}
\author{Juhee Hong}
\affiliation{\mbox{WCI Center for Fusion Theory, 
National Fusion Research Institute,} \\
Daejeon 305-806, Korea}
\author{P. H. Diamond}
\affiliation{\mbox{WCI Center for Fusion Theory, 
National Fusion Research Institute,}  \\
Daejeon 305-806, Korea} 
\affiliation{CASS and Department of Physics, University of California, \\
San Diego, La Jolla, 
CA 92093, United States}
\date{\today}

\begin{abstract}
The shear viscosity of the quark-gluon plasma is predicted to be 
lower than the 
collisional viscosity for weak coupling. The estimated ratio of the 
shear viscosity to entropy density is rather close to the ratio calculated 
by $\mathcal{N}=4$ super Yang-Mills theory for strong coupling, which 
indicates that the quark-gluon plasma might be strongly coupled. 
However, in presence of momentum anisotropy, the Weibel 
instability can arise and drive the turbulent transport. 
Shear viscosity can be lowered by enhanced 
collisionality due to turbulence, 
but the decorrelation time and its relation to underlying dynamics and 
color-magnetic fields have not been calculated self-consistently. 
In this paper, we use resonance broadening theory for strong turbulence 
to calculate the anomalous viscosity of the quark-gluon plasma for 
nonequilibrium. 
For saturated Weibel instability, we estimate the scalings of the 
decorrelation rate and viscosity and compare these with collisional 
transport. 
This calculation yields an explicit connection between the underlying 
momentum space anisotropy and the viscosity anomaly. 
\end{abstract}

\maketitle

\section{Introduction}

At sufficiently high temperature, transport in quark-gluon plasma can be 
described by weakly coupled theories. 
Given that typical particles have momentum $\sim T$, 
there are several important kinetic scales, in terms of weak coupling $g\ll 1$ 
\cite{Arnold}. 
First, static color-electric fields are screened at the Debye length 
$\sim 1/(gT)$. 
Second, (unlike traditional electromagnetic plasmas) static color-magnetic 
fields are confined at nonperturbative scales $\sim 1/(g^2T)$. 
Finally, dynamics is governed by particle collisions at macroscopic scales 
$\sim 1/(g^4T)$ where hydrodynamics can be applied. 
Transport in quark-gluon plasma has been studied primarily based on 
macroscopic hydrodynamics. 
However, there are mesoscopic scales, 
$1/T\ll \mbox{(distance)}\ll 1/(g^2T)$, 
where collective effects can be important and a 
magnetohydrodynamic description can be applied, as in electromagnetic plasmas. 

Transport coefficients have been calculated using the linearized Boltzmann 
equation \cite{AMYllog,AMYblog}. 
Taking account of two particle collisions, the ratio of the shear viscosity to 
entropy density is 
\st
\label{weakvis}
\frac{\eta_C}{s}\sim \frac{1}{g^4\ln(1/g)}  \, .
\stp
On the other hand, experimental data can be described by hydrodynamic 
simulations with an anomalously low viscosity. 
Comparing elliptic flow data with simulations, 
the shear viscosity of the quark-gluon plasma is deduced to be 
(see a review \cite{Teaney:2009qa}) 
\st
\label{etabys}
\frac{\eta}{s}\sim\frac{1\leftrightarrow 5}{4\pi}  \, .
\stp
Much thought has been devoted to the fact that the shear viscosity as deduced from 
data is much lower than the 
collisional viscosity \Eq{weakvis} when the coupling constant is small 
enough. 
\Eq{etabys} is rather close to the ratio predicted using 
$\mathcal{N}=4$ super Yang-Mills theory for strong coupling, 
$\eta/s=1/4\pi$ \cite{ads1,ads2}. 
One way to resolve the problem of why $\eta< \eta_C$ is indicated is to assume 
that the quark-gluon plasma is strongly coupled. 
Alternatively, instability effects have been suggested as a means for enhanced 
collisionality which can reduce $\eta$ \cite{Asakawa:2006tc,Asakawa:2006jn}. 
We will discuss this scenario in this work. 

When anisotropic momentum distributions occur, the Weibel\footnote{The 
Weibel instability arises in presence of momentum space anisotropy or 
temperature gradient \cite{Weibel}. It is cumulative effects of counter-streams and 
develops current filamentation.} 
instability can arise at soft momentum $\sim gT$. 
The Weibel instability of the quark-gluon plasma has been studied by 
transport theory, and, equivalently, in hard thermal loop dynamics 
\cite{mrow,mrowhtl,chromoweibel}. 
There have been numerical simulations and analytic studies of thermalization 
and cascade 
\cite{ALMbottom,Arnold:2004ih,ALMYth,Rebhan:2005re,Rebhan,AMYfate,AMk2,AMspec,Dumitru}. 
In electromagnetic plasmas, Weibel-excited random fields 
coherently scatter particles, and so reduce the rate of momentum transport 
\cite{Dupree1,Dupree2}. 
Similarly, turbulent color-magnetic fields might affect transport 
properties of the quark-gluon plasma. 
In that case, viscosity is not obtained solely by particle collisions, 
but instability effects must be also accounted for. 

Viscosity measures stress per velocity gradient. 
Since the stress tensor is $\sim T^4$ and the collision frequency is 
$\sim g^4T\ln(1/g)$ for soft momentum transfer, the collisional viscosity is 
$\eta_C\sim T^3/(g^4\ln(1/g))$, as in \Eq{weakvis}. 
In presence of instability driven fluctuations, we must take a possibly enhanced 
decorrelation frequency (due to interaction between particles and turbulent 
fields) into account when computing the transport. 
Although it depends on which mechanism (collisions or instabilities) is 
dominant for the relevant kinetic regime, the effective viscosity is roughly 
determined by 
\st
\eta\sim\frac{ \, \mbox{(Stress)} \, }{ \, \, \mbox{(Collision Frequency) 
+ (Decorrelation Frequency)} \, \, } \, .
\stp
In high temperature non-Abelian plasmas, instability arises at momentum $\lsim gT$. 
So, we guess that the decorrelation frequency is $\lsim gT$. 
Since the decorrelation frequency can be higher than the collision frequency, 
it follows that instability and momentum space scattering might lower the viscosity of the quark-gluon plasma 
\cite{Asakawa:2006tc,Asakawa:2006jn}. 

The actual quark-gluon plasma produced in relativistic heavy ion collisions 
is a complicated dynamic system. 
Calculating the viscosity requires us to understand the fluctuation 
dynamics and transport properties 
of the plasma in each stage. 
However, to investigate instability effects on viscosity, we consider a 
rather simple case in this work. 
According to numerical simulations, there is no significant difference 
between Abelian plasmas and non-Abelian plasmas in 1+1 dimensions: instability 
grows exponentially \cite{Arnold:2004ih,AMYfate,Rebhan:2005re}. 
Such Abelianization disappears in 3+1 dimensions, where 
instability growth is subexponential. 
To estimate the lower bound of the anomalous viscosity, we assume an 
Abelian regime in 1+1 dimensions which can be used to determine the maximum 
intensity of plasma instabilities and transport. 
In Section \ref{LinearInstability}, we briefly review the linear instability. 
We focus on the turbulent Weibel state for soft momentum $k\sim gT$. 
In Section \ref{qgp}, we analyze nonlinear particle-wave interaction using 
resonance broadening theory for strong turbulence. 
For saturated Weibel instability, 
we obtain the relation between the decorrelation frequency and turbulent 
color-magnetic fields. 
Following \cite{Abe1,Abe2} in electromagnetic plasmas, 
we calculate the decorrelation frequency and the 
anomalous viscosity of the quark-gluon plasma for nonequilibrium. 
Finally, we summarize our results in Section \ref{conclusion}.

\section{Linear Instability}
\label{LinearInstability}

In this section, we briefly review the linear analysis for the 
Weibel instability. 
We assume an Abelian regime by linearizing the equations of motion in the 
gauge field. 
In the next section, we consider nonlinear particle-wave interaction due 
to resonance broadening for strong turbulence. 

We linearize the distribution of hard particles as\footnote{For plasmas 
consisting of gluons, $f=2N_cf_g$, where $f_g$ is 
the distribution function of gluons per helicity and color. 
$\delta f=\delta f^aT^a$, where $\delta f^a$ and generators $T^a$ are 
in the adjoint representation. 
}
\st
f=\langle f\rangle +\delta f \, ,
\stp 
where $\langle f\rangle$ is color-neutral and anisotropic in momentum $\p$, 
and $\delta f$ is colored fluctuations. 
At mesoscopic scales, the kinetic equation of particles is the Vlasov 
equation
\st
v^\mu \partial_\mu \delta f^a + 
g(\bm{E}^a+\v\times\bm{B}^a)\cdot\frac{\partial 
\langle f\rangle}{\partial\p}=0 \, .
\stp 
where $v^\mu=p^\mu/E_\p$. 
Color-electromagnetic fields obey the non-Abelian Maxwell equation 
\st
\label{nonAbelMax}
\partial_\nu F^{\mu\nu,a}=J^{\mu,a}=
g\int\frac{d^3\p}{(2\pi)^3}v^\mu \delta f^a \, .
\stp

In Fourier space, the linear solution of the Vlasov equation is 
\st
\label{linsol}
\delta f^a(\omega,\k)
=-\frac{g(\bm{E}^a+\v\times\bm{B}^a)\cdot\frac{\partial \langle f\rangle}
{\partial \p}}{-i\omega+i\v\cdot\k} \, .
\stp
By plugging the solution to the non-Abelian Maxwell equation, we have 
\st
\label{eqstat}
ik_\nu F^{\mu\nu,a}=-g^2\int\frac{d^3\p}{(2\pi)^3}\frac{v^\mu (\bm{E}^a
+\v\times\bm{B}^a)\cdot\frac{\partial\langle f\rangle}{\partial\p}}
{-i\omega+i\v\cdot\k} \, .
\stp
This can be written as 
\st
\label{tensoreq}
\epsilon^{\mu\nu}A_\nu^a=0 \, ,
\stp
where we defined a tensor 
\st
\label{dielectrictensor}
\epsilon^{\mu\nu}\equiv (-\omega^2+k^2)g^{\mu\nu}-k^\mu k^\nu+\Pi^{\mu\nu} 
\stp
with the self-energy ($\epsilon$ is positive and infinitesimal)
\st
\label{selfE}
\Pi^{\mu\nu}=g^2\int\frac{d^3\p}{(2\pi)^3}\frac{\partial\langle f\rangle}
{\partial p^i}\left[-v^\mu g^{i\nu}+
\frac{v^\mu v^\nu k^i}{-\omega+\v\cdot\k-i\epsilon}\right] \, .
\stp
In the temporal gauge $A_0=0$, we have $\epsilon^{ij}E_j=0$, and 
the linear dispersion relation is 
\st
\label{lindisrel}
{\rm{det}}\, \epsilon^{ij}=0 \, .
\stp
Depending on the sign of $\Im \, \omega$, we have exponentially growing or 
damping solutions $\omega(\k)$. 
If there is an exponentially growing solution with $\Im \, \omega>0$, the 
quark-gluon plasma has instability that can drive turbulence.

\section{Nonlinear Particle-Wave Interaction}
\label{qgp}

In this section, we consider nonlinear particle-wave interaction due to 
resonance broadening. 
Resonance broadening theory is well defined for traditional electromagnetic 
plasmas (see Appendix \ref{emplasma}) and amounts to calculating phase space 
eddy diffusivity and its effects on particle trajectories\footnote{
These enter the linear response which determines the instability.}. 
We can apply resonance broadening theory to the relativistic 
quark-gluon plasma in momentum space. 
For strong turbulence, the linear dispersion relation can be 
extended to the nonlinear regime with a simple correction in the self-energy. 
For the Weibel instability at saturation, we calculate the diffusion 
coefficient (which is related to color-magnetic fields), 
the particle-wave decorrelation time, and the anomalous viscosity. 
The momentum space diffusion coefficient is determined by the saturation 
condition. 
This sets an effective root-mean-square turbulence intensity. 
This approach is made in the spirit of Prandtl's theory of pipe flow 
turbulence than of the familiar Kolmogorov cascade.

\subsection{Resonance Broadening}

The distribution function is written as 
\st
f=\langle f\rangle +f_{\omega,\k}+\tilde{f} \, , 
\stp
where $\langle f\rangle$ is the average over space, $f_{\omega,\k}$ is the 
coherent part with respect to color-electromagnetic fields, and $\tilde{f}$ 
represents 
fluctuations due to noise\footnote{$\tilde{f}$ is ignored in the quasilinear 
order.}. 
Taking the average over space, the mean field Vlasov equation becomes (see, 
for example, \cite{Nicholson}) 
\st
\label{qleq}
\frac{\partial}{\partial t}\langle f\rangle+g
\left\langle (\bm{E}^a+\v\times\bm{B}^a)\cdot\frac{\partial f_{\omega,\k}^a}
{\partial\p}\right\rangle=0 \, ,
\stp
where we used the fact that $f$ does not diverge at 
infinity\footnote{For spatially 
homogeneous $\langle f\rangle $, 
$\left\langle\frac{\partial f}{\partial x}\right\rangle
=\displaystyle \lim_{L\rightarrow\infty}\frac{1}{L}\int_{-L/2}^{L/2}dx \, 
\frac{\partial f}
{\partial x}=\lim_{L\rightarrow\infty}\frac{1}{L}
\left[f\left(x=\frac{L}{2}\right)-f\left(x=-\frac{L}{2}\right)\right]
=0$. 
In Section \ref{AnomVis}, the $\v\cdot\frac{\partial}{\partial\x}$ term will be 
revived in calculating the viscosity.}
 and $\langle \bm{E}^a\rangle=\langle\bm{B}^a\rangle=0$. 

Similar to the linear solution \Eq{linsol}, the coherent 
response $f_{\omega,\k}^a$ has a peak $\sim 1/(\omega-\v\cdot\k)$ 
corresponding to the resonance where particle velocity is equal to the phase 
velocity of color-electromagnetic waves. 
In presence of nonlinear interaction between particles and waves, the former 
are scattered by the ensemble of wave fields. 
As a result, the peak of the resonance is broadened 
(see, for example, \cite{PDbook}). 
To explain resonance broadening, we consider test particle dynamics in one 
dimension. 
In the linear order, the particle trajectory is assumed to be unperturbed, 
since nonlinear particle-wave interaction scatters the trajectory from the 
unperturbed one by $\delta x$. 
So, the coherent response is 
\st
f_{\omega,k}^a=-\int_0^\infty dt \, 
e^{i(\omega t-kx)+ik \, \delta x}
g(\bm{E}_{\omega,k}^a+\v\times\bm{B}_{\omega,k}^a)\cdot
\frac{\partial\langle f\rangle}{\partial\p} \, . 
\stp
By plugging this response to the quasilinear equation \Eq{qleq}, we obtain 
a diffusion equation \cite{Dupree1,Dupree2}
\st
\label{diffeq}
\left(\frac{\partial}{\partial t}
-\frac{\partial}{\partial\p}\cdot\bm{D}(\p)\cdot\frac{\partial}{\partial
\p}\right)\langle f\rangle=0 \, ,
\stp
where the diffusion tensor is given by the Lorentz force-force correlator 
with $\bm{F}_{\omega,k}^a=g(\bm{E}_{\omega,k}^a+\v\times\bm{B}_{\omega,k}^a)$ 
\st
\bm{D}(\p)=\int_{0}^\infty dt \, 
e^{i(\omega t-kx)+ik \, \delta x}
\langle \bm{F}_{\omega,k}^a \, \bm{F}_{\omega,k}^a
\rangle \, .
\stp
Since color-electromagnetic fields are turbulent, particles perform a random walk 
in momentum space. 
This diffusion scatters particles from their unperturbed trajectories, weakens 
the response, and eventually saturates the instability. 

The scatter of a trajectory can be calculated by taking the average over the 
probability density function (pdf). 
We assume that $\delta p$ has a Gaussian pdf 
\st
\label{dppdf}
\mbox{pdf}\, [\delta p]=\frac{1}{\sqrt{\pi D t}} \, 
e^{-\frac{(\delta p)^2}{Dt}} \, .
\stp
Performing the Gaussian integral, we have 
\bg
\langle e^{i(\omega t-kx)+ik \, \delta x} \rangle_{\rm{pdf}}
&=&
\int
\frac{ d \, (\delta p)}{\sqrt{\pi Dt}} 
e^{-\frac{(\delta p)^2}{Dt}} 
e^{i(\omega t-kx)+ik\int dt \, (\delta v)}
\, , \nonumber\\
&\simeq&e^{i(\omega -vk)t-\frac{k^2Dt^3}{4\bar{E}_\p^2}} \, ,
\nd
where we approximated $\int dt \, (\delta v)\simeq t \, (\delta p)/E_\p$ 
\footnote{Since $\v=\p/E_\p$, 
$\delta v=(\delta p)/E_\p-p \, (\delta E_\p)/E_\p^2$, 
where we ignore $\delta E_\p$ with a diffusive pdf of \Eq{dppdf}. 
Assuming a similar Gaussian pdf of $\delta E_\p$, we obtain a consistent 
$t^3$ factor of resonance broadening. 
} 
and replaced $E_\p$ by the averaged $\bar{E}_\p\equiv(\int d^3\p \, E_\p
\langle f\rangle)/(\int d^3\p \, \langle f\rangle)$. 
From the coefficient of $t^3$ term, we define the 
particle-wave decorrelation time $t_c$ 
\st
\label{diffc}
\left(\frac{1}{t_c}\right)^3\equiv \frac{k^2D}{4\bar{E}_\p^2} \, .
\stp
Here, $t_c$ is the time scale it takes the wave ensemble to scatter a particle 
by wavelength $\sim 1/k$ from its unperturbed trajectory. 

\begin{figure}
\includegraphics[width=0.45\textwidth]{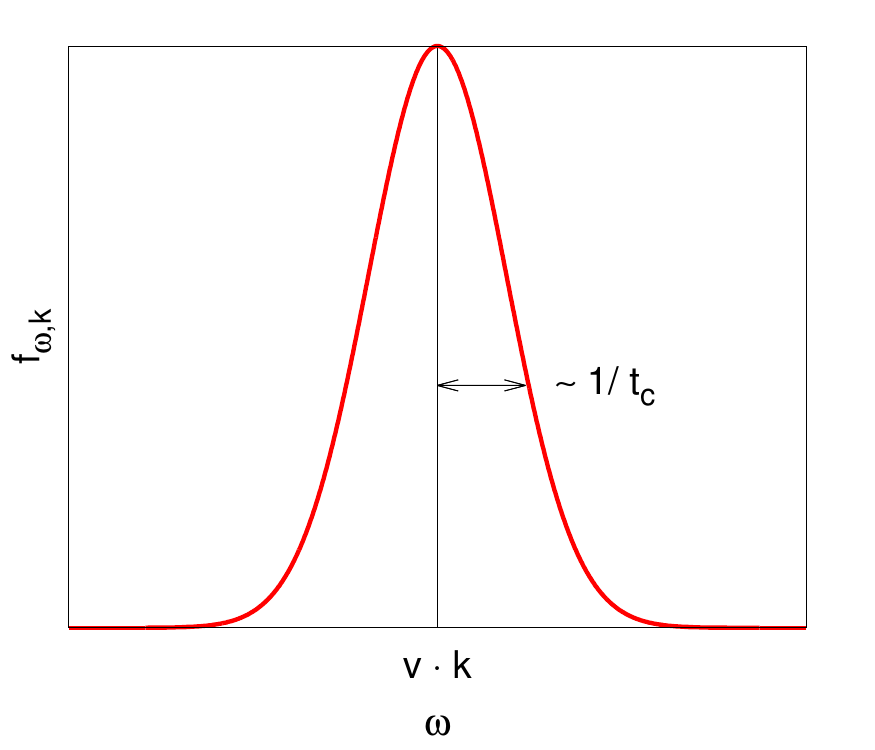}
\caption{(Color online) 
The coherent response $f_{\omega,\k}$ has a resonance at $\omega=\v\cdot\k$. 
Due to nonlinear particle-wave interaction, the resonance peak of a delta 
function $\delta(\omega-\v\cdot\k)$ is broadened with a width proportional 
to the decorrelation rate $1/t_c$.}
\label{rbt}
\end{figure}

The principal effect of nonlinear particle-wave interaction is to broaden the 
resonance peak of a delta function to a resonance with a width 
proportional to the decorrelation rate $1/t_c$. 
Thus, we can use the Lorentzian approximation for strong turbulence 
as an approximation (see \Fig{rbt}) 
\st
\label{Lorentz}
\int dt \, e^{i(\omega-\v\cdot\k)t-t^3/t_c^3}
\simeq
\frac{i}{\omega-\v\cdot\k+i/t_c} \, .
\stp
In this regard, 
within resonance broadening theory for strong turbulence, the self-energy 
\Eq{selfE} acquires a nonlinear correction which amounts 
to the replacement $\omega\rightarrow \omega+i/t_c$.

\subsection{Diffusion Coefficient}

In the absence of static color-electromagnetic fields, the diffusion tensor due 
to color-magnetic excitations is\footnote{In this work, we consider 
nonlinear particle-wave interaction for saturated Weibel instability. So, $\omega$ and $\k$ in the 
summation satisfy the linear dispersion relation.} \cite{Dupree3}
\st
\label{diffcoeffdef}
D=
\sum_{\omega,\k}
(g\v\times\delta\bm{B}_{\omega,\k}^a)
\frac{i}{\omega-\v\cdot\k+i/t_c}
(g\v\times\delta\bm{B}_{\omega,\k}^a) \, ,
\stp
where we used the Lorentzian approximation \Eq{Lorentz}. 
For most unstable modes, the wave vector is along the direction of anisotropy 
and color-magnetic excitations is perpendicular to the direction\footnote{
Color-electric excitations are 
$\delta \bm{E}_{\omega,\k}^a=\delta E_{\omega,k}^a\hat{\x}$. 
Since color-electric fields are related to color-magnetic fields by the 
non-Abelian Maxwell equation, we consider only color-magnetic fields.} 
\st
\label{1dpropagate}
\k=k\hat{\z} \qquad \mbox{and} \qquad
\delta\bm{B}_{\omega,\k}^a=\delta B_{\omega,k}^a \, \hat{\y} \, .
\stp
Since the Weibel instability is purely growing, 
we set $\omega=i\gamma$, where $\gamma$ is the growth rate. 
Then the diffusion coefficient is 
\st
D= 
\sum_{\omega,k}
g^2v_x^2|\delta B_{\omega,k}^a|^2
\frac{1}{\gamma+1/t_c+iv_zk} \, .
\stp

We now consider how large color-magnetic excitations can grow. 
When the Weibel instability saturates, color-magnetic excitations stop 
growing ($\gamma=0$). 
So, we have 
\st
D=\sum_{\omega,k}
g^2v_x^2|\delta B_{\omega,k}^a|^2
\frac{1/t_c}
{(1/t_c)^2+(v_zk)^2} \, ,
\stp
where the imaginary part vanished because it is an odd function of $k$. 
It can be simplified for ``strong turbulence'' where the particle-wave 
decorrelation time $t_c$ is so short compared to the time scale 
$\sim 1/(\v\cdot\k)$ that the condition $(1/t_c)^2\gg (v_zk)^2$ is 
satisfied. 
Ignoring $(v_zk)^2$ in the denominator, we obtain 
\st
\label{diffcoeff}
D\simeq
\sum_{\omega,k}
g^2v_T^2|\delta B_{\omega,k}^a|^2
\frac{1}{1/t_c} \, ,
\stp
where we replaced $v_x^2$ by the thermal velocity $v_T^2$ 
\footnote{The thermal velocity squared $v_T^2\sim 1$ is a typical 
velocity squared of particles in the quark-gluon plasma.}. 
With the definition of the decorrelation time \Eq{diffc}, we determine the 
relation between the decorrelation time and the intensity of 
color-magnetic excitations at saturation, namely
\st
\label{reltcB}
\left(\frac{1}{t_c}\right)^4
\simeq
\frac{k^2}{4\bar{E}_\p^2}
\sum_{\omega',k'}
g^2v_{T}^2|\delta B_{\omega',k'}^a|^2 
\, .
\stp
Here, $t_c$ gives the time scale for scattering of a particle, that is, 
the trajectory mixing time.

\subsection{Decorrelation Time}

The particle-wave decorrelation time can be determined from the nonlinear 
dispersion relation. 
As discussed below \Eq{Lorentz}, the self-energy has a nonlinear correction 
due to the resonance broadening 
\st
\label{PiNL}
\Pi_{\rm{NL}}^{ij}=
g^2\int\frac{d^3\p}{(2\pi)^3}\frac{\partial\langle f\rangle}
{\partial p^l}\left[-v^i g^{lj}+
\frac{v^i v^j k^l}{-\omega+\v\cdot\k-i/t_c}\right] \, .
\stp
Following \cite{Romatschke:2003ms}, given an isotropic distribution 
$\langle f(\p^2)\rangle_{\rm{iso}}$, we make an anisotropic distribution 
by the rescaling of the $\hat{\z}$ direction 
\st
\langle f\rangle=\langle f(\p^2+\xi p_z^2)\rangle_{\rm{iso}} \, .
\stp
Here, $\xi>-1$ is the anisotropy parameter: 
$-1<\xi<0$ corresponds to a stretch and $\xi>0$ corresponds to a squeeze in 
the $\hat{\z}$ direction. 
By a change of variables to $\tilde{p}\equiv p\sqrt{1+\xi v_z^2}$, \Eq{PiNL} 
can be calculated as: 
\st
\Pi_{\rm{NL}}^{ij}=
m_D^2\int\frac{d\Omega}{4\pi}\frac{v^i}{(1+\xi v_z^2)^2}
\left[v^j+\xi v_z \hat{z}^j+\frac{(\xi+1)v^jv_zk}
{\omega-v_zk+i/t_c}
\right] \, ,
\stp
where 
\st
m_D^2=-\frac{g^2}{2\pi^2}\int_0^\infty dp\, p^2
\frac{d\langle f\rangle_{\rm{iso}} }{dp} \, .
\stp 

\begin{figure}
\includegraphics[width=0.45\textwidth]{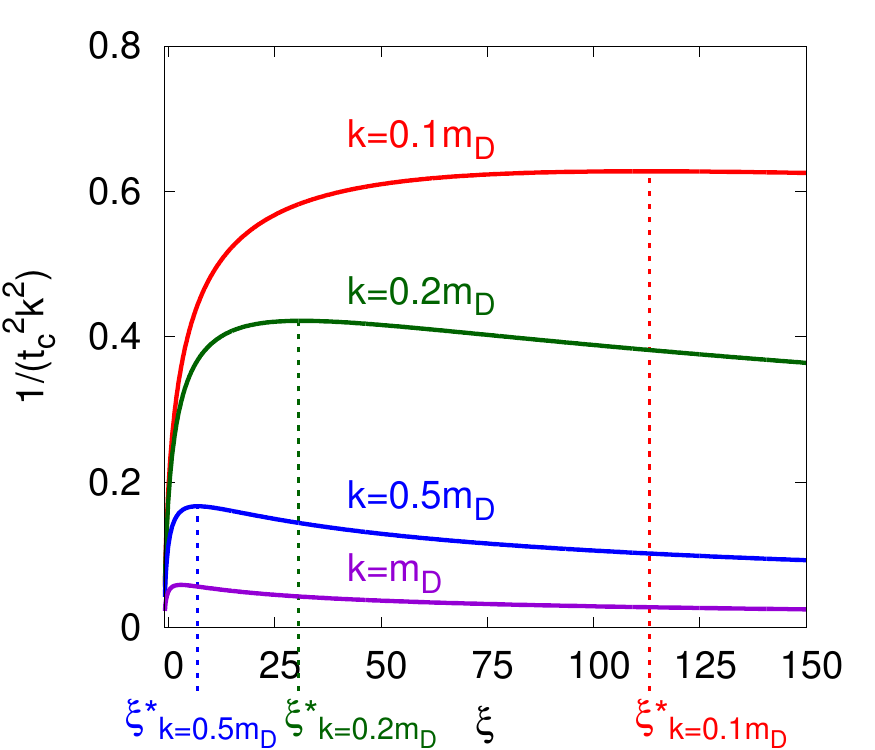}
\caption{(Color online) 
As anisotropy grows, the decorrelation time decreases until $\xi_k^*>0$ 
for the wave vector $k$. 
For strong turbulence, the anisotropic parameter must be in the regime where 
$1/(t_c^2k^2)\gg v_z^2$. 
Since $v_z^2 \ll 1$, $\xi \approx \xi^*_{k}$ most likely satisfies the 
condition. 
For low $k$, this regime corresponds to an extreme squeeze in the momentum 
$\hat{\z}$ direction of an initially isotropic distribution. 
}
\label{validxi}
\end{figure}

In the case of \Eq{1dpropagate}, the dispersion relation is 
\st
\label{nldisrel}
-\omega^2+k^2+\Pi_{\rm{NL}}^{xx}=0 \, .
\stp
For strong turbulence, when the Weibel instability saturates, 
the self-energy term is 
\st
\Pi_{\rm{NL}}^{xx}
\simeq\frac{m_D^2}{4}\left[\frac{1}{\xi}+\frac{(\xi-1)}{\xi}\frac{
\arctan\sqrt{\xi}}{\sqrt{\xi}}\right] 
-\frac{m_D^2t_c^2k^2}{4}
\left[-\frac{3(\xi+1)}{\xi^2}
+\frac{(\xi+1)(\xi+3)}{\xi^2}\frac{\arctan\sqrt{\xi}}{\sqrt{\xi}}\right]
\, .
\stp
From \Eq{nldisrel}, we determine the decorrelation time 
\st
\label{tcsol}
t_c^2\simeq
\frac{k^2+\frac{m_D^2}{4}\Big[\frac{1}{\xi}+\frac{(\xi-1)}{\xi}
\frac{\arctan\sqrt{\xi}}{\sqrt{\xi}}\Big]}
{\frac{m_D^2}{4}\Big[-\frac{3(\xi+1)}{\xi^2}+\frac{(\xi+1)(\xi+3)}{\xi^2}
\frac{\arctan\sqrt{\xi}}{\sqrt{\xi}}\Big]k^2}  
\qquad
\mbox{for strong turbulence}
\, ,
\stp
where functions of $\xi$ in the square brackets are positive. 
Since the decorrelation time is taken to be short for strong turbulence, 
it must satisfy 
\st
\frac{1}{t_c^2k^2}\simeq
\frac
{\frac{m_D^2}{4}\left[-\frac{3(\xi+1)}{\xi^2}+\frac{(\xi+1)(\xi+3)}{\xi^2}
\frac{\arctan\sqrt{\xi}}{\sqrt{\xi}}\right]}  
{k^2+\frac{m_D^2}{4}\left[\frac{1}{\xi}+\frac{(\xi-1)}{\xi}
\frac{\arctan\sqrt{\xi}}{\sqrt{\xi}}\right]}
\gg v_z^2 \, ,
\stp
which gives the validity regime for the anisotropy parameter $\xi$ 
(see \Fig{validxi}). 
As anisotropy grows, the decorrelation time decreases until $\xi_k^*>0$ for 
the wave vector $k$. 
Noting $v_z^2\ll 1$, $\xi$ around $\xi_k^*$ most likely satisfies the strong 
turbulence condition. 
For low $k$, this regime corresponds to an extreme squeeze in the momentum 
$\hat{\z}$ direction of an initially isotropic distribution. 
This might apply to the early stage of relativistic heavy ion collisions. 

At soft momentum $k\sim gT$, the scale of the decorrelation time \Eq{tcsol} is 
\st
\label{tcorder}
t_c\sim \frac{1}{k} \, .
\stp
Using Eqs. (\ref{reltcB}) and (\ref{tcsol}), we determine the saturation 
level of color-magnetic excitations 
\st
\frac{1}{4\bar{E}_\p^2}\sum_{\omega,k}g^2v_T^2|\delta B_{\omega,k}^a|^2
\simeq
\left[\frac{\frac{m_D^2}{4}\Big[-\frac{3(\xi+1)}{\xi^2}+\frac{(\xi+1)(\xi+3)}{\xi^2}
\frac{\arctan\sqrt{\xi}}{\sqrt{\xi}}\Big]}  
{k^2+\frac{m_D^2}{4}\Big[\frac{1}{\xi}+\frac{(\xi-1)}{\xi}
\frac{\arctan\sqrt{\xi}}{\sqrt{\xi}}\Big]}\right]^2
k^2 
\qquad\mbox{for strong turbulence} 
\, .
\stp
Thus, the scale of the saturated color-magnetic field is\footnote{This scale 
corresponds to when the covariant derivative 
($D=\partial-igA \, \sim \, i(p-gA)$) 
cannot be treated perturbatively, $A\sim\frac{E_\p}{g}$ 
\cite{Arnold:2004ih,AMYfate}.}
\st
\label{dBorder}
\delta B_{\omega,k} \sim \frac{k E_\p}{g} \, . 
\stp

\subsection{Anomalous Viscosity}
\label{AnomVis} 

In this section, we follow the strategy in \cite{Abe2} to calculate 
the anomalous viscosity. 
This is a somewhat artificial way to obtain viscosity, but it allows us to 
estimate its basic scalings. 
We assume $\langle f\rangle$ is spatially inhomogeneous. 
For simplicity, we make $v_x$ depend on $x$ by 
the replacement 
\st
v_x \, \rightarrow \, v_x- \frac{\tilde{p}^2}{p^2}u(x) \, , 
\stp 
where $\tilde{p}^2=p^2(1+\xi v_z^2)$ and 
$u(x)$ is the mean flow\footnote{We work in the local rest frame, 
$u(x)=0$.}. 
Then we take a second moment ($2p_x^2-p_y^2-p_z^2$) of the 
diffusion equation \Eq{diffeq}. 
The corresponding energy-momentum tensor is 
\bg
\label{2ndmom}
2T^{xx}-T^{yy}-T^{zz}&=&
\int \frac{d^3\p}{(2\pi)^3}\frac{2p_x^2-p_y^2-p_z^2}{E_\p}\langle f\rangle 
\, ,\nonumber\\
&=&\frac{1}{(2\pi)^3}
\int d\Omega\frac{2v_x^2-v_y^2-v_z^2}{(1+\xi v_z^2)^2} 
\, 
\int_0^\infty dp\, p^3 \langle f\rangle_{\rm{iso}}
\, .
\nd
From the coefficient of velocity gradient in the corresponding tensor, 
we determine the viscosity 
\st
\eta_A
=\frac{2T^{xx}-T^{yy}-T^{zz}}{-4\frac{\partial u}{\partial x}} 
\, .
\stp
 
In the case of \Eq{1dpropagate}, the diffusion equation is 
\st
\label{diffeq2}
\left(\frac{\partial}{\partial t}+\v\cdot\frac{\partial}{\partial \x}\right)
\langle f\rangle 
\simeq\sum_{\omega,k}g^2v_T^2|\delta B_{\omega,k}^a|^2\frac{1}{1/t_c}
\frac{\partial^2 \langle f\rangle}{\partial p_z^2} \, ,
\stp
where we used the diffusion coefficient \Eq{diffcoeff} for strong turbulence. 
For inhomogeneous $\langle f\rangle$, we revived the 
$\v\cdot\frac{\partial}{\partial\x}$ term 
\st
\v\cdot\frac{\partial\langle f\rangle}{\partial \x}
\simeq -v_T^2\tilde{p}\frac{d \langle f\rangle}{d \tilde{p}}
\frac{\partial u}{\partial x} \, ,
\stp
where we replaced $v_x^2$ by $v_T^2$. 
Taking a moment $(2p_x^2-p_y^2-p_z^2)$, the left hand side in \Eq{diffeq2} 
becomes 
\st
\label{lhs}
\mbox{(LHS)}=
\frac{\partial}{\partial t}(2T^{xx}-T^{yy}-T^{zz}) 
+
\frac{v_T^2}{(2\pi)^3}\int d\Omega\frac{2v_x^2-v_y^2-v_z^2}
{(1+\xi v_z^2)^2}
\left[-\int_0^\infty dp \, p^4\frac{d\langle f\rangle_{\rm{iso}}}{dp}\right]
\frac{\partial u}{\partial x}
 \, .
\stp
The right hand side is approximated as follows. 
First, we take a derivative 
\st
\frac{\partial^2\langle f\rangle}{\partial p_z^2}=
\frac{(\xi+1)^2 p_z^2}{\tilde{p}^2}
\frac{d^2\langle f\rangle}{d\tilde{p}^2}
+\frac{(\xi+1)}{\tilde{p}}\frac{d\langle f\rangle}{d\tilde{p}}
-\frac{(\xi+1)^2p_z^2}{\tilde{p}^3}
\frac{d\langle f\rangle}{d\tilde{p}}
\, ,
\stp
where only the second term contributes to the viscosity. 
Second, we take a moment $(2p_x^2-p_y^2-p_z^2)$ 
\st
\mbox{(RHS)}=(\xi+1)
\sum_{\omega,k}g^2v_T^2|\delta B_{
\omega,k}^a|^2\frac{1}{1/t_c}
\int\frac{d^3\p}{(2\pi)^3}
\frac{2p_x^2-p_y^2-p_z^2}{E_\p}
\frac{1}{\tilde{p}}
\frac{d\langle f\rangle}{d\tilde{p}}
\, .
\stp
Finally, we compare with \Eq{2ndmom} for $(2T^{xx}-T^{yy}-T^{zz})$ to write 
\st
\label{rhs}
\mbox{(RHS)}
=
-(\xi+1)\sum_{\omega,k}g^2v_T^2|\delta B_{\omega,k}^a|^2\frac{1}{1/t_c}
\frac{  
 \Big[-\int_0^\infty dp \, p^2 \frac{d\langle f\rangle_{\rm{iso}}}{dp} \Big] }
{  \Big[\int_0^\infty dp \, p^3\langle f\rangle_{\rm{iso}}\Big]  }
\left(2T^{xx}-T^{yy}-T^{zz}\right) 
\, .
\stp
For static state where $\frac{\partial}{\partial t}(2T^{xx}-T^{yy}-T^{zz})=0$, 
we equate \Eq{lhs} to \Eq{rhs} to determine the anomalous viscosity 
\st
\eta_A
\simeq
\frac{\frac{1}{2(4\pi)^2}\Big[\frac{\xi+3}{\xi(\xi+1)^2}+
\frac{(\xi-3)}{\xi(\xi+1)}\frac{\arctan\sqrt{\xi}}{\sqrt{\xi}}\Big]
\Big[-\int_0^\infty dp \, p^4\frac{d\langle f\rangle_{\rm{iso}}}{dp}\Big]
\Big[\int_0^\infty dp \, p^3\langle f\rangle_{\rm{iso}}\Big]}
{\displaystyle
\sum_{\omega,k}g^2|\delta B_{\omega,k}^a|^2\frac{1}{1/t_c}
\Big[-\int_0^\infty dp \, p^2\frac{d\langle f\rangle_{\rm{iso}}}{dp}\Big]
} 
\, ,
\stp
where 
the function of $\xi$ in the square brackets is positive for $\xi \approx 
\xi_{k}^*$ in \Fig{validxi}. 
As excited color-magnetic field intensity increases due to instability growth, 
the anomalous viscosity decreases because the effective collision frequency 
increases. 
Since we determined the saturation level of color-magnetic fields \Eq{reltcB} 
and the decorrelation time \Eq{tcsol}, the anomalous viscosity is 
given by 
\begin{multline}
\eta_A
\simeq
\frac{v_T^2}{2(8\pi)^2}\left[\frac{\xi+3}{\xi(\xi+1)^2}+
\frac{(\xi-3)}{\xi(\xi+1)}\frac{\arctan\sqrt{\xi}}{\sqrt{\xi}}\right]
\\
\times
\left[
\frac{k^2+\frac{m_D^2}{4}\Big[\frac{1}{\xi}+\frac{(\xi-1)}{\xi}
\frac{\arctan\sqrt{\xi}}{\sqrt{\xi}}\Big]}
{\frac{m_D^2}{4}\Big[-\frac{3(\xi+1)}{\xi^2}+\frac{(\xi+1)(\xi+3)}{\xi^2}
\frac{\arctan\sqrt{\xi}}{\sqrt{\xi}}\Big]}  
\right]^{3/2} 
\frac{
\Big[-\int_0^\infty dp \, p^4\frac{d\langle f\rangle_{\rm{iso}}}{dp}\Big]
\Big[\int_0^\infty dp \, p^3\langle f\rangle_{\rm{iso}}\Big]}
{\Big[-\int_0^\infty dp \, p^2\frac{d\langle f\rangle_{\rm{iso}}}{dp}\Big]
\bar{E}_\p^2 |k|}
\\ 
\Big. \Big. \mbox{for strong turbulence} \, . \quad
\end{multline}
We expect that this gives the lower bound of the anomalous viscosity 
in presence of the maximum intensity of the Weibel instability. 

Similar to the thermal velocity in nonrelativistic electromagnetic plasmas, we 
define the ``thermal momentum'' as:
\bg
\label{thermalp}
p_T^2&\equiv&\int\frac{d^3\p}{(2\pi)^3}
\frac{p_x^2}{E_\p}
\langle f\rangle
\nonumber\\ 
&=&\frac{1}{8 \pi^2}\left[
\frac{1}{\xi}+\frac{(\xi-1)}{\xi}\frac{\arctan\sqrt{\xi}}{\sqrt{\xi}}
\right]\int_0^\infty dp \, p^3\langle f\rangle_{\rm{iso}} \, ,
\nd
where the function of $\xi$ in the square brackets is positive. 
Then using \Eq{tcorder} for soft momentum $k\sim gT$, the scaling trend of the anomalous viscosity is\footnote{
The anisotropy parameter $\xi$ is a constant in this work. 
} 
\st
\label{etaAorder}
\eta_A\sim \frac{p_T^2}{1/t_c} \, ,
\stp
where $t_c$ at saturation is given by \Eq{tcsol}. 
This corresponds to \Eq{emetascale}, in that viscosity is roughly the ratio of 
the thermal velocity squared to the decorrelation frequency in electromagnetic 
plasmas. 
We note that $1/t_c$ sets the effective collision frequency.

\section{Summary and Discussions}
\label{conclusion}

In this work, we used resonance broadening theory for strong turbulence, 
$(1/t_c)^2\gg (\v\cdot\k)^2$, to analyze nonlinear particle-wave interaction 
in the quark-gluon plasma. 
To determine the maximum intensity of plasma instabilities and transport, we 
assumed an Abelian regime in 1+1 dimensions.
With the wave vector along the anisotropy axis, the saturation level of 
color-magnetic excitations is 
\st
k^2\sum_{\omega',k'}
|\delta B_{\omega',k'}^a|^2 
\simeq
\frac{4\bar{E}_\p^2}{g^2v_T^2}
\left(\frac{1}{t_c}\right)^4
\, ,
\stp
where $t_c$ gives the time scale for scattering of a particle. 
For saturated Weibel instability, we calculated the particle-wave 
decorrelation time and the anomalous viscosity 
\bg
t_c^2&\simeq&
\frac{k^2+\frac{m_D^2}{4}\Big[\frac{1}{\xi}+\frac{(\xi-1)}{\xi}
\frac{\arctan\sqrt{\xi}}{\sqrt{\xi}}\Big]}
{\frac{m_D^2}{4}\Big[-\frac{3(\xi+1)}{\xi^2}+\frac{(\xi+1)(\xi+3)}{\xi^2}
\frac{\arctan\sqrt{\xi}}{\sqrt{\xi}}\Big]k^2}  
\, ,
\\
&&\left.\right.
\nonumber
\\
\eta_A
&\simeq&
\frac{v_T^2}{2(8\pi)^2}\left[\frac{\xi+3}{\xi(\xi+1)^2}+
\frac{(\xi-3)}{\xi(\xi+1)}\frac{\arctan\sqrt{\xi}}{\sqrt{\xi}}\right]
\nonumber\\
&&\times
\left[
\frac{k^2+\frac{m_D^2}{4}\Big[\frac{1}{\xi}+\frac{(\xi-1)}{\xi}
\frac{\arctan\sqrt{\xi}}{\sqrt{\xi}}\Big]}
{\frac{m_D^2}{4}\Big[-\frac{3(\xi+1)}{\xi^2}+\frac{(\xi+1)(\xi+3)}{\xi^2}
\frac{\arctan\sqrt{\xi}}{\sqrt{\xi}}\Big]}  
\right]^{3/2} 
\frac{
\Big[-\int_0^\infty dp \, p^4\frac{d\langle f\rangle_{\rm{iso}}}{dp}\Big]
\Big[\int_0^\infty dp \, p^3\langle f\rangle_{\rm{iso}}\Big]}
{\Big[-\int_0^\infty dp \, p^2\frac{d\langle f\rangle_{\rm{iso}}}{dp}\Big]
\bar{E}_\p^2 |k|}
\, .
\quad
\label{lowerboundeta}
\nd
Here, the anisotropy parameter is $\xi \approx \xi^*_k$ in 
\Fig{validxi}, which corresponds to an extreme squeeze in the momentum 
$\hat{\z}$ direction of an initially isotropic distribution. 
We expect that \Eq{lowerboundeta} gives the lower bound of the anomalous 
viscosity in presence of the maximum intensity of the Weibel instability.
At soft momentum $k\sim gT$, the typical scales 
of the color-magnetic fields, the decorrelation time, 
and the anomalous viscosity are, respectively: 
\st
\delta B \sim T^2 \, , \qquad 
t_c \sim \frac{1}{gT} \, , 
\qquad  \mbox{and}\qquad
\eta_A \sim \frac{T^3}{g} \, . 
\stp
We note that the scale of the anomalous viscosity at $k\sim gT$ is much lower than the 
leading order collisional viscosity $\eta_C\sim T^3/g^4$. 

As discussed in Introduction and \Eq{etaAorder}, 
the effective viscosity is given by stress per effective collision frequency, 
so 
\st
\eta\sim \frac{p_T^2}{\, 1/t_{coll}+1/t_{c} \, } \, ,
\stp
where $1/t_{coll}$ is the collision frequency and $1/t_c$ is the decorrelation 
frequency. 
Although it depends on the relevant kinetic regime, 
the scale of the decorrelation frequency $1/t_c$ at $k\sim gT$ 
is much higher than the collision frequency 
$1/t_{coll}\sim g^4T$. 
As compared to the collisional viscosity $\eta_C\sim T^3/g^4$, the effective 
viscosity thus can be lowered to $\eta\sim T^3/g$ due to enhanced collisionality by 
nonlinear particle-wave interaction. 
This indicates that instability effects can be dominant in certain stages 
of quark-gluon plasma transport. 

We focused on strong turbulence to consider nonlinear and stochastic 
particle-wave interaction due to resonance broadening. 
In addition to particle-wave interaction, there are other nonlinear effects 
(including wave-wave interaction) which might be important in non-Abelian  
plasmas. 
Numerical simulations indicate that gluon self-interactions might 
control the saturation of the Weibel instability in 3+1 dimensions 
\cite{AMYfate,Rebhan:2005re}. 
However, there are limitations to simulations and their interpretation. 
Therefore, analytic study of nonlinear theory is 
essential to extract information from the simulations and to understand 
thermalization of the quark-gluon plasma. 
We hope to discuss a systematic nonlinear analysis on the quark-gluon plasma 
instabilities in future papers.

\bigskip
\bigskip

\noindent{\bf ACKNOWLEDGMENTS}

We would like to thank G. M. Fuller for useful discussions. 
This work is supported by the World Class Institute (WCI) Program of 
the National Research Foundation (NRF) funded by the Ministry of 
Science, ICT and Future Planning (MSIP) of Korea (WCI 2009-001).

\appendix

\section{Electromagnetic Plasmas}
\label{emplasma}

In this appendix, we discuss the Weibel instability in traditional 
electromagnetic plasmas by using resonance broadening theory 
\cite{Abe1,Abe2}. 
We consider plasmas consisting of electrons, ignoring motions of 
heavier ions. 
The analysis parallels to Section \ref{qgp} except 
\begin{itemize}
\item The coupling constant $g$ (or plasmon mass $m_D/\sqrt{3}$) is replaced 
by the electric charge $e$ (or plasma frequency 
$\omega_p=\sqrt{4\pi ne^2/m}$, where $n$ is the number density and 
$m$ is the mass of electrons). 
\item The phase space is velocity $\v$ space instead of momentum $\p$ space. 
\end{itemize}

Transport can be described by a diffusion equation \cite{Dupree1,Dupree2} 
\st
\label{emdiffeq}
\left(\frac{\partial}{\partial t}
-\frac{\partial}{\partial\v}\cdot\bm{D}(\v)\cdot\frac{\partial}{\partial
\v}\right)\langle f\rangle=0 \, ,
\stp
where the diffusion tensor is given by the Lorentz force-force correlator with 
$\bm{F}=e(\bm{E}+\v\times\bm{B})/m$. 
Assuming $\delta v$ has a Gaussian probability density function 
\st
\mbox{pdf}\, [\delta v]
=
\frac{1}{\sqrt{\pi D t}}e^{-\frac{(\delta v)^2}{Dt}} \, ,
\stp
the pdf average with change of a trajectory is 
\bg
\langle e^{i(\omega t-kx)+ik\delta x} \rangle_{\rm{pdf}} 
&=&
\int\frac{d(\delta v)}{\sqrt{\pi Dt}}e^{
-\frac{(\delta v)^2}{Dt}}e^{i(\omega t-kx)+ik\int dt (\delta v)}
\, ,
\nonumber\\
&\simeq&
e^{i(\omega -vk)t-\frac{k^2Dt^3}{4}} \, .
\nd
The particle-wave decorrelation time is defined as 
\st
\label{emtcDrel}
\left(\frac{1}{t_c}\right)^3\equiv \frac{k^2D}{4} \, .
\stp

By the Lorentzian approximation \Eq{Lorentz} for strong turbulence, 
the diffusion coefficient due to magnetic excitations is 
\st
D=
\sum_{\omega,\k}
\left(\frac{e}{m}\v\times\delta\bm{B}_{\omega,\k}\right)
\frac{i}{\omega-\v\cdot\k+i/t_c}
\left(\frac{e}{m}\v\times\delta\bm{B}_{\omega,\k}\right) \, .
\stp
For simplicity, we consider one-dimensional propagation of plasmas, 
\Eq{1dpropagate}. 
When the Weibel instability saturates, the diffusion coefficient is 
\st
D\simeq
\sum_{\omega,k}
\frac{\omega_p^2}{4\pi nm}v_T^2
|\delta B_{\omega,k}|^2
\frac{1}{1/t_c} \, ,
\stp
where we replaced $v_x^2$ by $v_T^2$. 
Using \Eq{emtcDrel}, the saturation level of magnetic excitations is 
\st
\label{emsat}
\left(\frac{1}{t_c}\right)^4
\simeq
\frac{k^2}{4}
\sum_{\omega',k'}
\frac{\omega_p^2}{4\pi nm}v_{T}^2|\delta B_{\omega',k'}|^2 
\, .
\stp

In resonance broadening theory, the nonlinear dispersion relation is 
given by 
\st
\label{disrelem}
\omega^2+\omega_p^2\int d^3\v\left[
v_x\frac{\partial \langle f\rangle}{\partial v_x}
+\frac{kv_x^2}{\omega-v_zk+i/t_c}
\frac{\partial \langle f\rangle}
{\partial v_z}
\right]
=k^2 \, .
\stp
With an anisotropy parameter $\xi>-1$, 
electrons obey the Maxwellian distribution 
\st
\langle f\rangle =\frac{\sqrt{\xi+1}}{(\sqrt{2\pi}v_T)^3}e^{-\frac{v^2+\xi
v_z^2}{2v_T^2}} \, , 
\stp
where $v_T$ is the thermal velocity\footnote{We normalized the distribution, 
$\int d^3\v \langle f\rangle = 1 $. 
The thermal velocity is the averaged velocity, 
$\int d^3\v \, v_x^2 \langle f\rangle = v_T^2$. 
}. 
At saturation, the particle-wave decorrelation time is 
\st
\label{disrelem2}
t_c^2\simeq
\frac{1}{v_T^2}\left(\frac{1}{k^2}+\frac{1}{\omega_p^2}\right)
\qquad\mbox{for strong turbulence}
\, .
\stp
Since we used the strong turbulence approximation, $(1/t_c)^2\gg (v_zk)^2$, 
it must satisfy 
\st
\label{emvalidxi}
\frac{1}{k^2/\omega_p^2+1}\gg \frac{v_z^2}{v_T^2}
\, .
\stp
Noting $v_z^2\ll v_T^2$, this condition is valid for $k^2\lsim \omega_p^2$. 
Using \Eq{emsat}, the saturation level of magnetic excitations is 
\st
\frac{1}{16\pi nm}\sum_{\omega,k}|\delta B_{\omega,k}|^2
\simeq
\frac{\omega_p^2v_T^2k^2}{(k^2+\omega_p^2)^2}
\qquad\mbox{for strong turbulence}
\, .
\stp

To calculate the anomalous viscosity, we assume that the Maxwellian 
distribution 
depends on space by the replacement $v_x\rightarrow v_x-u(x)$. 
The diffusion equation is 
\st
\left(\frac{\partial}{\partial t}+\v\cdot\frac{\partial}{\partial \x}\right)
\langle f\rangle 
\simeq
\sum_{\omega,k}\frac{\omega_p^2}{4\pi nm}v_T^2|\delta B_{\omega,k}|^2
\frac{1}{1/t_c}
\frac{\partial^2\langle f\rangle}{\partial v_z^2} \, ,
\stp
where we revived the $\v\cdot\frac{\partial}{\partial\x}$ term for 
inhomogeneous $\langle f\rangle$. 
Taking a second moment $(2v_x^2-v_y^2-v_z^2)$ on both sides, we obtain 
\st
\frac{\partial}{\partial t}(2T_{EM}^{xx}-T_{EM}^{yy}-T_{EM}^{zz})
+\frac{nmv_T^2\xi}{(\xi+1)}\frac{\partial u}{\partial x}
\simeq -(\xi+1)
\sum_{\omega,k}\frac{\omega_p^2}{4\pi nm}|\delta B_{\omega,k}|^2
\frac{1}{1/t_c}
(2T_{EM}^{xx}-T_{EM}^{yy}-T_{EM}^{zz}) 
\, ,
\stp
where the corresponding stress tensor is 
\st
2T_{EM}^{xx}-T_{EM}^{yy}-T_{EM}^{zz}
=nm\int d^3\v \, (2v_x^2-v_y^2-v_z^2) \, \langle f\rangle  \, .
\stp
For static state where $\frac{\partial}{\partial t}(2T_{EM}^{xx}-T_{EM}^{yy}
-T_{EM}^{zz})=0$, the anomalous viscosity is determined as 
\st
\eta_A
\simeq\frac{\frac{ nmv_T^2\xi}{4(\xi+1)^2}}{
\displaystyle
\sum_{\omega,k}\frac{\omega_p^2}{4\pi nm}|\delta B_{\omega,k}|^2
\frac{1}{1/t_c}
} \, .
\stp
As magnetic field intensity increases, the anomalous viscosity decreases. 
Since the saturation level of magnetic fields and the decorrelation time are 
determined by Eqs. (\ref{emsat}) and (\ref{disrelem2}), we have 
\st
\eta_A\simeq\frac{nmv_T\xi k^2}{16(\xi+1)^2} 
\left(
\frac{1}{k^2}+\frac{1}{\omega_p^2}
\right)^{3/2}
\qquad\mbox{for strong turbulence}\, .
\stp
For $k^2\lsim \omega_p^2$, the scaling trend of the anomalous viscosity is 
given by 
\st
\label{emetascale}
\eta_A\sim
\frac{v_T^2}{1/t_c} \, ,
\stp
where we used $v_Tk\sim 1/t_c$.

\bibliography{viscosity}

\end{document}